\documentclass[conference]{IEEEtran}%
\usepackage{amsfonts}
\usepackage{amssymb}
\usepackage{amsmath}
\usepackage{graphicx}
\usepackage{color}%
\usepackage{url}
\providecommand{\U}[1]{\protect\rule{.1in}{.1in}}

\newtheorem{definition}{Definition}

\begin{document}

\title{Ultra-Reliable Communication in \\ 5G Wireless Systems}
\author{
\authorblockN{Petar Popovski}\\
\authorblockA{Department of Electronic Systems, Aalborg University\\
Email: petarp@es.aau.dk}}
\maketitle

\begin{abstract}
Wireless 5G systems will not only be ``4G, but faster''. One of the novel features discussed in relation to 5G is \emph{Ultra-Reliable Communication (URC)}, an operation mode not present in today's wireless systems. URC refers to provision of certain level of communication service almost $100$ \% of the time. Example URC applications include reliable cloud connectivity, critical connections for industrial automation and reliable wireless coordination among vehicles. This paper puts forward a systematic view on URC in 5G wireless systems. 
It starts by analyzing the fundamental mechanisms that constitute a wireless connection and concludes that one of the key steps towards enabling URC is revision of the methods for encoding control information (metadata) and data. It introduces the key concept of \emph{Reliable Service Composition}, where a service is designed to adapt its requirements to the level of reliability that can be attained. The problem of URC is analyzed across two different dimensions. 
The \emph{first dimension} is the type of \emph{URC problem} that is defined based on the time frame used to measure the reliability of the packet transmission. Two types of URC problems are identified: long-term URC (URC-L) and short-term URC (URC-S). The \emph{second dimension} is represented by the type of \emph{reliability impairment} that can affect the communication reliability in a given scenario. The main objective of this paper is to create the context for defining and solving the new engineering problems posed by URC in 5G.
\end{abstract}

\section{Introduction}
\subsection{5G Wireless and its Operating Regions}

Cellular wireless systems from 2G to todayÕs 4G have been evolving towards offering the users connectivity at increasingly higher data rates. While this trend is expected to continue in the fifth generation (5G) wireless systems, there are strong indications \cite{Petar5G,METIS} that 5G will not only be ``4G, but faster'', but will also feature at least two new operating modes: 
\begin{itemize}
\item \emph{Ultra-Reliable Communication (URC)}: This is an operation mode not present in today's cellular wireless systems and refers to provision of certain level of communication service almost $100$ \% of the time.
\item \emph{Massive M2M (Machine-to-Machine) Communication (MMC)}: This mode already emerges as an extension of the 4G LTE systems and refers to support of a massive number (tens of thousands) machines in a given area. 
\end{itemize}

\begin{figure}
	\centering
		\includegraphics[width=8.5cm]{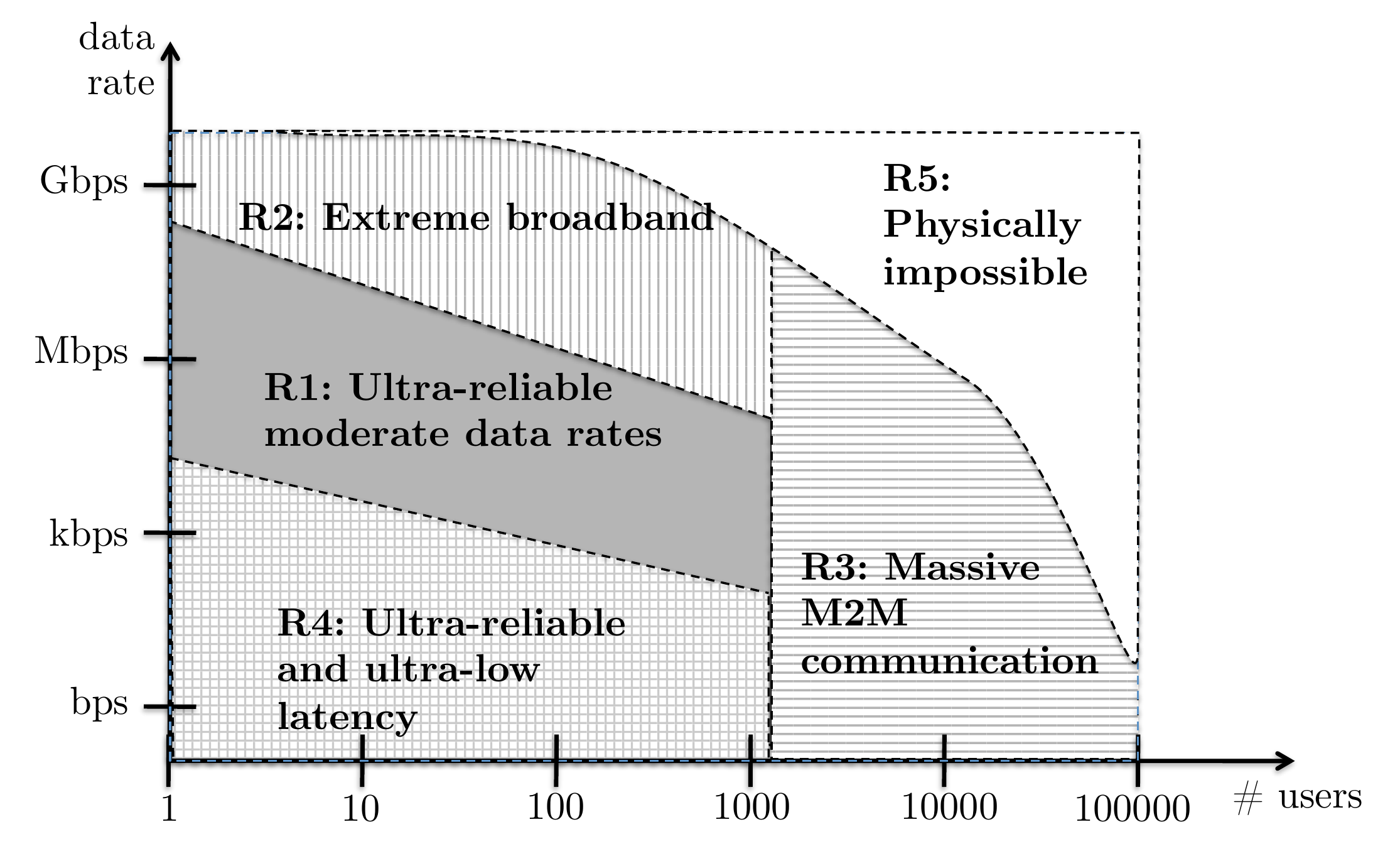}
		\caption{Operating regions of the 5G wireless systems.}
		\label{fig:5Gregions}
\end{figure}

Fig.~\ref{fig:5Gregions} illustrates the expected operating regions of 5G wireless systems defined in the context of the data rate vs. the number of connected devices in a service area. The numbers are not precise and only  depict the order of magnitude. At present, the large and diverse ecosystem of wireless systems is dominated by cellular technologies for wide-area use, such as 4G LTE (Long Term Evolution) and local high-speed use, such as Wi-Fi. These systems operate in the region \textbf{R1}, whose shape outlines that the data rate of each user decreases as the user population increases. Clearly, 5G wireless will support the same operating region; however, the rates of \textbf{R1} will be rather \emph{moderate} in the context of 5G, considering that there will be also extreme data rates, see region \textbf{R2}. However, differently from today's systems, the rates in \textbf{R1} in 5G will be, for some services, supported in an \emph{ultra-reliable} manner. For example, the data rate of $50$ Mbps will be offered with very high reliability ($>99 \%$) or strict latency guarantees, which is not the case today. The region \textbf{R2} features extreme broadband rates and it is very often mistakenly referred to as ``the 5G wireless'' due to the very active research agenda that contributes to this region, including: $60$ GHz spectrum use, massive MIMO, full duplex wireless, etc. Contrary to the broadband regime, the region \textbf{R3} and most of \textbf{R4} feature \emph{lowband}\footnote{We are not using the obvious term \emph{narrowband} as it would refer to the data rates of the digital systems in the beginning of $90$s.} data rates. In lowband communication, the messages sent from/to the devices are short. In the region \textbf{R3}, these short messages are coming from a large number of machines/sensors in e.g. the smart grid or environmental sensing. In the region \textbf{R4} the short messages are exchanged with very low latency, as in e.g. traffic-safety-related communication among vehicles or critical industrial control. The operation in \textbf{R5} is impossible due to fundamental physical and information-theoretic limits.

\subsection{What is URC? Motivating Examples.}
\label{sec:URCexamples}

Despite the large proliferation, commercial wireless technologies have not attained the stage in which connectivity is guaranteed almost $100$\% of the time. The reason is that the commercial wireless technologies are designed to offer relatively good connectivity most of the time, but offer almost zero data rate in areas with poor coverage, under excessive interference or when the network resources are overloaded. On the other hand, wireless technology continues to enter into new application areas and an increasing number of services will start to depend critically on the availability of wireless links that offer at least minimal communication quality.
The term ``commercial'' is emphasized to differentiate from wireless systems used by the military or law enforcement agencies, where URC is achieved under a completely different set of techno-economic constraints and dedicated spectrum allocation. 

Referring to Fig.~\ref{fig:5Gregions}, URC is relevant for more than one region, as illustrated by the following examples. 
\begin{itemize}
\item \emph{Reliable cloud connectivity}. All cloud-based services assume that Internet connectivity is available during the large percentage of the time. For mobile devices, as the wireless connectivity becomes more available and reliable, the cloud services will be reshaped in order to rely even more on the wireless connection. One could ask, for example: how to design a cloud application knowing that $99.9$ \% of the time there is at least 1 Mbps available and $99$ \% of the time there is at least 50 Mbps available? The reliability can also refer to guaranteed low latency for transferring a message of a given size, which is an enabler of the ``Tactile Internet'' \cite{Tactile}. This type of URC is featured in the region \textbf{R1}.    
\item \emph{Vehicle-to-Vehicle (V2V) wireless coordination.} In a futuristic scenario, the cars will be wirelessly interconnected in a very reliable way, such that there is no need to use traffic lights at a crossing, the cars will coordinate through short wireless messages. Enabling such a high reliability requires fundamentally new transmission techniques and access protools for sending short wireless messages. This type of URC is illustrative for the region \textbf{R4}. 
\item \emph{Alarm from a massive set of sensors} 5G wireless will enable deployment of large-scale distributed cyber-physical systems for e.g. smart grid or industrial control. These require lowband communications and most of the time the short messages are of low importance or redundant (e.g. sensor reporting correlated measurements). However, in some cases there can be a critical event (e.g. a protective relay in smart grid) that needs to be reported with very high reliability. The challenge is how to support critical operation that coexists with the usual lowband traffic. This type of URC is situated in region \textbf{R3} of Fig.~\ref{fig:5Gregions}.
\end{itemize}

These three examples reflect today's perspective on wireless services and are therefore limited in depicting the scope of URC. Already in 3G there were claims for connectivity ``anywhere and anytime''; nevertheless, it is hard to perceive this claim beyond its marketing value, since no cellular operator is willing to guarantee a  data rate to an individual device $>99$\% of the time. However, once the ultra-reliable feature is available, then one can talk about wireless as a commodity that is truly available ``anywhere and anytime'' and it is hard to foresee all the applications that will be built assuming the existence of such wireless links.

\section{Elements of Ultra-Reliable Communication}

\subsection{Anatomy of a Digital Data Connection}
\label{sec:AnatomyData}
In order to understand the design needs for URC, we need to go back to the fundamental constituents of a 
digital data connection. Assume that Alice wants to send data to Bob over a Additive White Gaussian Noise (AWGN) channel with bandwidth $W$ and SNR of $\gamma$. The classical result in information theory states that the maximal data rate at which Alice can send to Bob is the \emph{channel capacity} \cite{Shannon48}: 
\begin{equation}\label{eq:CapacityGaussian}
C(W,\gamma)=W \log_2 (1+\gamma) \quad \mathrm{[bps]}
\end{equation}
The practical interpretation is that one needs to send a very large volume of data over a very large number of symbols in order to use the data rate given by (\ref{eq:CapacityGaussian}) and guarantee that Bob decodes the data with almost zero probability of error. 

However, what is rarely discussed in relation to (\ref{eq:CapacityGaussian}) is the role of the \emph{control data} or the \emph{metadata} that is a pre-condition to carry any data communication. 
In order to see the impact of the metadata, let us assume that Alice transmits to Bob using $n$ \emph{channel uses}. A channel use is the smallest, atomic unit of communication that can be sent from the transmitter to the receiver. Let one channel use take $T_s$ seconds. In AWGN, a channel use is represented by a complex baseband symbol to which a complex Gaussian noise is added when arriving at the receiver. Achieving the capacity in~(\ref{eq:CapacityGaussian}) requires that one single codeword spreads over infinitely many  channel uses. However, for all practical purposes it can be assumed that the formula is valid when $n$ is very large, with a remark that the probability of error is $p_{e,d}$, a value close to zero such that instead of capacity, we can speak of a throughput $C(W,\gamma)(1-p_{e,d})$. The total amount of data sent by Alice during the $n$ channel uses is $D$, then the relationship is 
\begin{equation}
D=n \cdot T_s \cdot C(W,\gamma) \qquad \mathrm{[bits]}
\end{equation}
The formula (\ref{eq:CapacityGaussian}) assumes that Bob is in a state where he knows he receives data from Alice. To achieve this state, Alice uses $m$ channel uses preceding the $n$ data channel uses to send metadata, also called \emph{header}. The header is a short packet that has its own integrity (CRC) check and carries $H$ bits of data. It is usually $H \ll D$ and the data rate $R_H$ of the header is chosen to be very low, since the reception of the header is a condition to receive the data. Let $R_H$ be chosen  such that the probability of error in receiving the header is $p_{e,h}$. The effective \emph{goodput} from Alice to Bob achieved is:
\begin{equation} \label{eq:GAB}
G_{AB}=\frac{D}{(m+n)T_s} (1-p_{e,h})(1-p_{e,d}) \qquad \mathrm{[bps]}
\end{equation}
Here $(1-p_{e,h})(1-p_{e,d})$ is the probability that Bob receives the data correctly, since it is mandatory that at first he receives the header. Note that we have not included the requirement that Bob sends backs ACK message to Alice and she receives it correctly; that would only further decrease the throughput. 

The high-speed wireless systems, such as LTE, put a major focus on how to efficiently transmit \emph{large data volume} i.e. $D \gg H$ and $n \gg m$. In that case the following two features can be used: (1) large data means that one can use methods (codes, modulation) that are almost capacity-achieving. (2) the size of metadata is small compared to the size of data, such that even if the metadata is sent suboptimally (e. g. repetition coding and very low $R_H$), its overall effect on the system performance is negligible. Since $H \ll D$, then even with very low rate $R_H$, the value of $m$ can be neglected within the goodput expression~(\ref{eq:GAB}). The low value of $R_H$ is used to guarantee that $p_{e,h} \ll p_{e,d}$, such that $(1-p_{e,h}) \approx 1$ in (\ref{eq:GAB}). 

In URC, the objective is to make $(1-p_{e,h})(1-p_{e,d})$ very high and thus satisfy the high reliability levels. 
One idea could be that we do not use inefficient decoding for the header and instead combine the header and the data in a single packet and encode them efficiently. This would be a packet that spans over $n+m$ channel uses and with probability of error $q_{e,d}$ where $(1-q_{e,d}) > (1-p_{e,h})(1-p_{e,d})$. The problem with such a transmission is that Bob needs to know \emph{a priori} that he should decode the transmission. To see this, consider the case where there are two possible receivers of Alice's message, Bob and Carol and Alice sends a packet to Bob. If the data and metadata are jointly encoded, then both Bob and Carol must decode everything and only after decoding, Bob decides to accept the data for himself, while Carol drops it. Clearly, for Carol this is not efficient in terms of energy, but it is the price to be paid to have an improved transmission reliability. This type of metadata/data encoding is an example of the tradeoff between energy efficiency and very high reliability. 

Separate encoding of header and data becomes even worse when the data packets are short, such that metadata and data are roughly of the same size, $H \approx D$. In that case $m$ becomes comparable to $n$, even larger if the coding for the header is done in an inefficient way in order to increase the robustness. As a result, the goodput in (\ref{eq:GAB}) decreases. In this situation the joint encoding of metadata and data becomes even more relevant, since the overall data size that needs to be encoded increases to $H+D$. 
The recent fundamental work on rates/error probabilities for finite block length~\cite{Polyanskiy} indicates that with packets of short size, say with $H=80$ and $D=128$ it is more efficient to encode a data block of size $H+D=208$ bits. 

We make a slight digression to relate our discussion of data-metadata encoding to the case of analog communication systems. Why is analog voice communication considered to be very robust and treated as the ``last resort'' in many critical systems, such as airplane or military? Analog voice communication is inherently suitable for graceful degradation: as the communication conditions worsens the voice quality decreases, but is still comprehensible. To interpret in terms of data and metadata, one can say that the data is the content of the speech that is transmitted, while the metadata is the information about the speaker. The metadata is sent \emph{continuously} as the analog voice contains biometric features that identify the speaker. It can be concluded that the robustness of the analog voice communication is rooted in the fact of joint encoding of metadata and data instead of sending the metadata only at the beginning and then supplement it with data. 

The main message of this discussion is that URC requires reconsideration of the traditional ways that are used to send metadata and data. New transmission methods should consider fully or partially joint encoding of data and metadata, along with the optimization of the associated tradeoffs, notably the tradeoff between energy efficiency and reliability. 

\subsection{Reliable Service Composition}

Our working definition of reliability is:
\begin{definition}
Reliability is the probability that a certain amount of data from one peer is successfully transmitted to another peer within a given deadline or time frame.
\end{definition}

Ultimately, a communication system should support reliable transfer of data for a service/application that resides in the higher protocol layers. All the other procedures are only auxiliary building blocks to support the main goal. The reliability requirements (latency, data rate, error probability) at the higher layers can, in principle, be translated into reliability requirements to each of the lower layers. However, this is   putting conservative requirements to the lower layers, as the following example shows. Consider a cloud computing service, where the requirement is that 
the user has the perception that the computing/memory resources are local and this is translated into latency requirement of e.g. $0.5$ seconds. However, this number does not specify the amount of data transferred during that time, such that one needs to account for the highest amount of data possible. However, adjusting the system only to the highest data volume will lead to prohibitively high rate requirements that are very difficult to satisfy: Either the system has to pre-reserve resources that are idle most of the time or the service needs to accept certain degradation compared to what had been originally requested. The second option is a viable solution to keep high system efficiency while providing a high level of reliability.

A reliability requirement, such as \emph{``transfer of data packets that have at most B bytes with a delay D less than L seconds in $99$\% of the attempts''} creates a rather simple criterion to see whether the system meets the requirement or not. However, it is important to ask \emph{Does the service need to fail whenever the reliability requirement is not met?} In order to answer ÒnoÒ to this question, we need to reconsider the way in which a certain communication service is composed. \emph{Reliable service composition (RSC)} is a way to specify different versions of a service, such that when the communication conditions are worsened, the Quality of Experience(QoE) gracefully degrades to the service version that can be reliably supported, instead of having a binary decision ``service available/not available''. The concept of graceful degradation of a service is not new, see for example scalable video coding~\cite{ScalableVideoCoding}. However, video and its perception naturally allows for graceful degradation; in RSC, the objective is to \emph{design} services that offer certain level of functionality when it is not possible to get the full one.

\begin{figure}
	\centering
		\includegraphics[width=7cm]{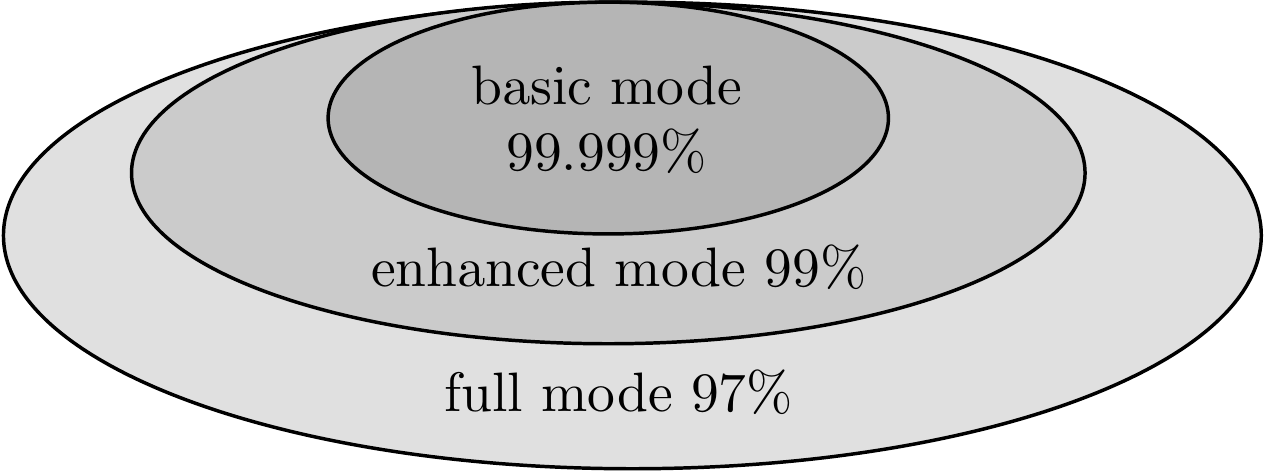}
		\caption{Illustration of the Reliable Service Composition (RSC).}
		\label{fig:RSC}
\vspace{-12pt}
\end{figure}

Fig.~\ref{fig:RSC} depicts the main idea behind RSC (the percentages are only provisional). Let us consider RSC in the case of Vehicle-to-Vehicle (V2V) communication.  
The basic version of the service is available $99.999$ \% of the time. In the V2V setting, the basic version could involve transmission of a small set of warning/safety messages without certification. The fact that the set of messages transferable in the basic mode is limited can be used to design efficient low-rate mechanisms to transfer those messages. An enhanced version of the service is available $99$ \% of the time, includes limited certification and guarantees for transfer of payload of size $D_1$ within time $T_1$ with probability $99.9$ \%. The full version is available $97$ \% of the time, includes full certification and  guarantees for transfer of payload of size $D_2 >D_1$ within time $T_2 < T_1$ with probability $99.9$ \%. The key issue in making RSC operational is to have reliable criteria to detect in which version the system should apply at a given time. The design of data/metadata for each service version should be integrated in an overall protocol that can flexibly switch between modes as the dynamic conditions dictate. 

\section{Types of URC Problems}

The variability of the requirements across the three URC examples in Section~\ref{sec:URCexamples}, indicates that there are different classes of URC problems. In this section we use the latency parameter as a dimension across which we identify two different types of URC problems:
\begin{itemize}
\item \emph{URC over a long term (URC-L)}: This type of URC deals with problems that require minimal rate over a longer period ($>10$ ms), such as minimal rate for a connection to a public cloud in a densely populated area, etc. 
\item \emph{URC in a short term (URC-S)}: Problems with very stringent latency requirements ($\leq 10$ ms), such as vehicles communicating at a crossroad, teleprotection in smart grid, etc. 
\end{itemize}
The $10$ ms value comes from the METIS project \cite{METISD62}. It should be noted that a specific class of URC for emergency communications falls under the umbrella of URC-L. The objective in URC for emergency is to provide minimal connectivity when the infrastructure is damaged or non-existent. It does involve aspects of radio access, which is the main theme in this paper, but it also involves techniques from ad hoc networking, delay-tolerant networking and self-healing, which reside in the higher layer of the protocol stack and are outside the scope of this paper.

\subsection{URC over a Long Term (URC-L)}

\begin{figure}[t]
	\centering
		\includegraphics[width=8.5cm]{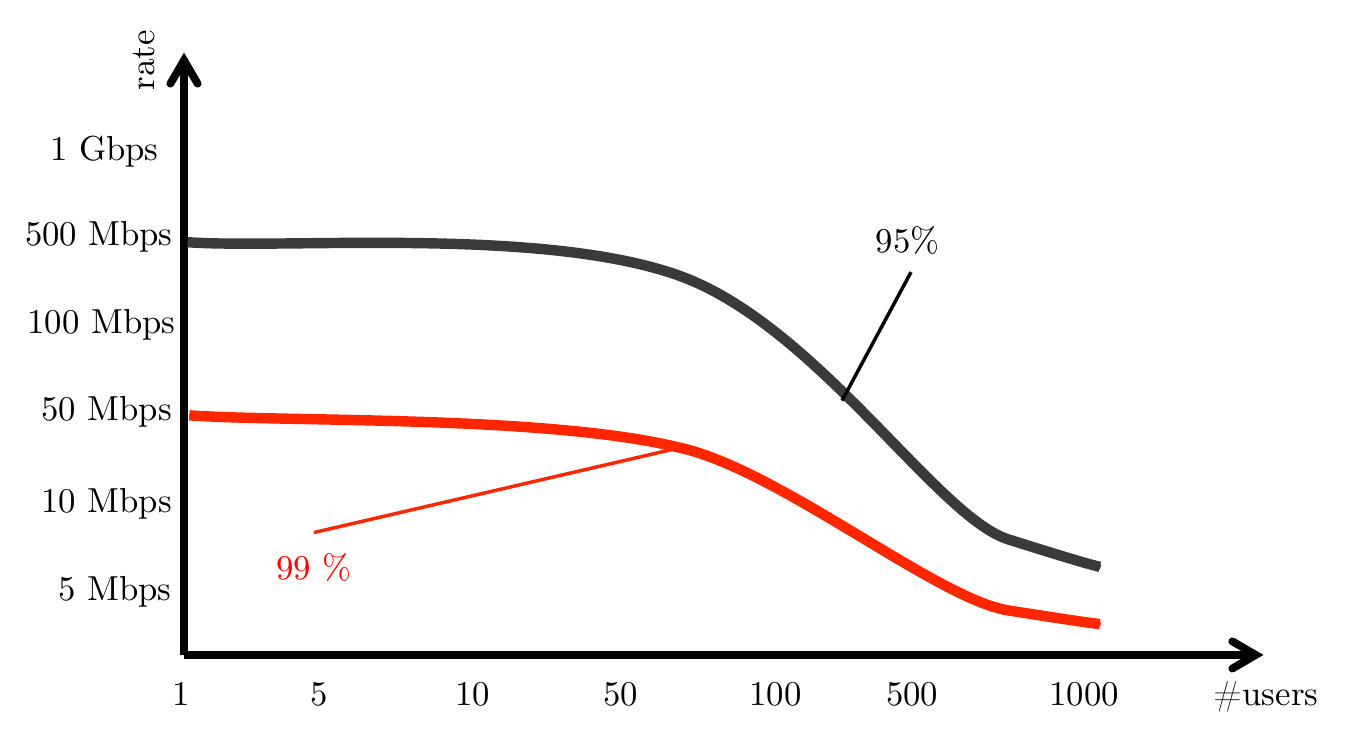}
		\caption{Illustrative graph for URC-L where the average rate is depicted as a function of the number of users that share resources.}
		\label{fig:URCLnrusers}
\end{figure}

The general problem in URC-L is how to guarantee rates, with high probability, to one or multiple users over longer periods. For example, in reliable cloud connectivity, an operator would like to guarantee to the user a certain connectivity level within a given coverage area. Here we define the coverage area as the area in which a user is able to receive control information from the infrastructure during $99$ \% of the time. We provide two illustrative examples of target performance requirements for URC-L:
\begin{itemize}
\item When the user has a dedicated communication resource, then in the coverage area he should be offered at least $500$ Mbps during $95$ \% of the time and at least $50$ Mbps during $99$ \% of the time. 
\item When the user needs to share the resources with multiple users, then the target performance is depicted on Fig.~\ref{fig:URCLnrusers}.  
\end{itemize}
In both cases the average rate is calculated over a time window $T_W$ larger than $10$ ms, for example $T_W=1$ second. Fig.~\ref{fig:URCLnrusers} suggests that, as long as the number of users is up to $50$, then we can put forward the requirements for user with a dedicated resource. As the number of users grows beyond $50$ and becomes massive, then less resources remain for each user and the rate should degrade gracefully. 

Supporting URC-L can rely on using known techniques, but optimized in a 
new setup and new target performance figures. For example, Massive MIMO \cite{Marzetta} is an emerging technology that is a good candidate to support the requirements of URC-L. Massive MIMO operates with many spatial degrees of freedom and it could be used either to achieve extremely high reliability in supporting a given user (the first requirement above) or efficiently multiplex many users (Fig.~\ref{fig:URCLnrusers}). 
 
\subsection{URC over a Short Term (URC-S)}

\begin{figure}[t]
	\centering
		\includegraphics[width=8.5cm]{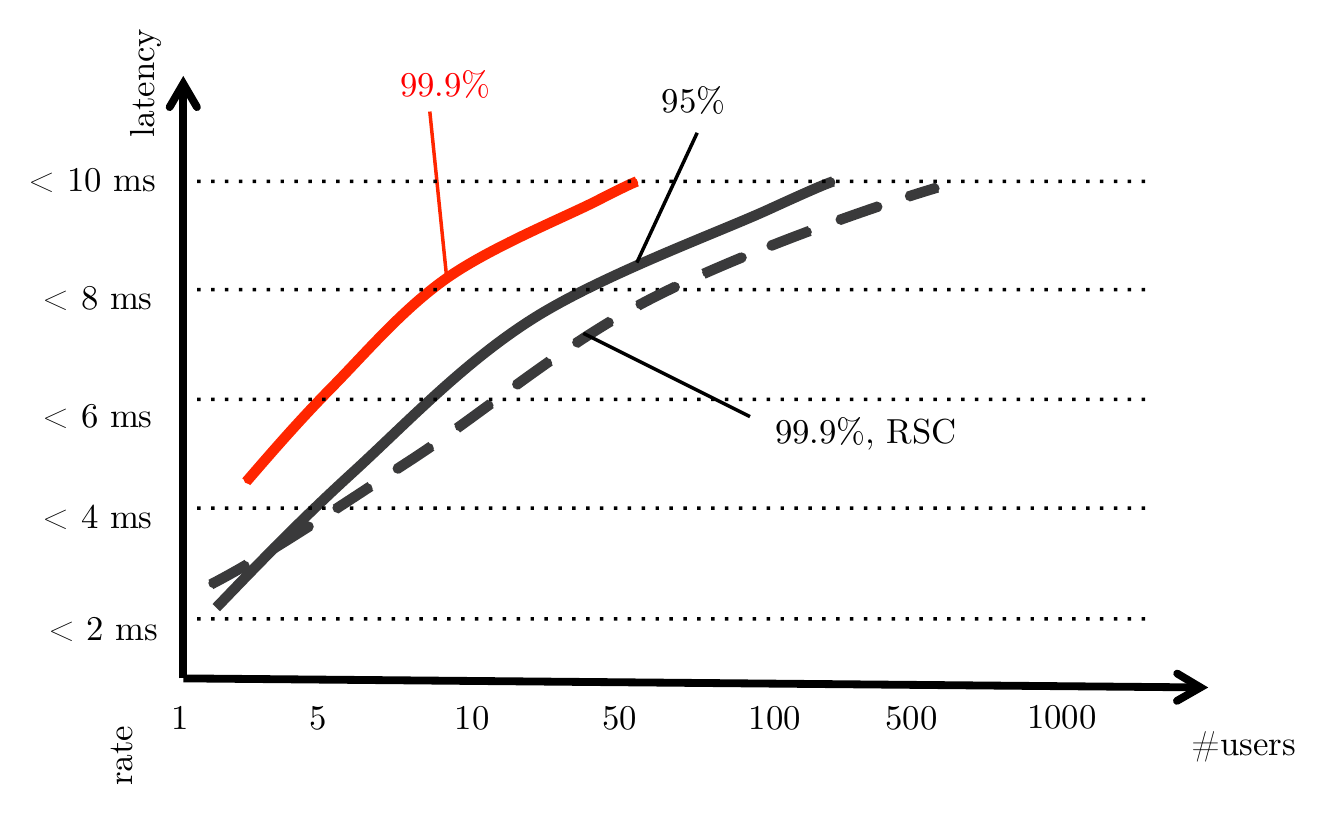}
		\caption{Illustrative performance requirements for URC-S where the latency is depicted as a function of the number of users.}
		\label{fig:URCSnrusers}
\end{figure}

In the case of URC-S, the focus is on how to deliver a certain portion of data under a very stringent latency requirement. Similar to URC-L, here we could also consider the latency for a single user that has dedicated resources or multiple users that need to satisfy latency requirements by sharing the resources. When there are multiple users, a significant part of the latency budget may be consumed due to the competition among the users (e.g. collisions in ALOHA-like protocols). An illustration of the target latency requirements with multiple competing users is given on Fig.~\ref{fig:URCSnrusers}. The full lines depict possible requirements for the performance in terms of latency/reliability when the service requirements are fixed and the messages have size of at most $D$ bits. If the service is created with Reliable Service Composition (RSC), then the dashed curve depicts the latency requirements when the basic mode of RSC is considered. In basic mode, each user sends at most $D_b$ bits, where $D_b<D$. This illustrates the fact that, when each user has a small set of possible messages, then an efficient design of data and metadata can lead to protocols with significantly optimized latency performance. 

There is a methodological difference between URC-L and URC-S in the following sense: while URC-L can rely on the bounds and the coding methods related to classical information theory, where the codeword length is very large, URC-S should rely on the techniques for coding short packets as well as the fundamental results from the area of coding for finite blocklength~\cite{Polyanskiy}. We illustrate how these results can be used to design systems with guaranteed reliability.  Let us fix the target packet error probability to $\epsilon$ and assume that there is an AWGN channel with SNR of $\gamma$. Let $n$ be the number of channel uses over which the codeword should span. The following relation is given in \cite{Polyanskiy}:
\begin{equation} \label{eq:FIniteChannelUses}
\log_2 M(n,\epsilon) \approx nC - \sqrt{nV} Q^{-1}(\epsilon)+\frac{1}{2} \log_2 n
\end{equation}
For given $n$ and $\epsilon$, $M(n,\epsilon)$ is the maximal number of different messages that can be sent over the $n$ channel uses such that the probability of reception error is $\epsilon$. $C$ is the capacity of AWGN channel for given $\gamma$, while $V$ is the dispersion of the channel, also dependent on the SNR and defined in \cite{Polyanskiy}. The function $Q$ is the standard function $Q(x)=\int_{x}^{\infty} \frac{1}{\sqrt{2 \pi}} e^{-t^2/2} \mathrm{d}t$. We would like to put a different perspective on (\ref{eq:FIniteChannelUses}). Let us assume that a message of size $10$ bytes needs to be sent over a point-to-point channel; this corresponds to $80$ bits, such that the total number of possible messages is $M=2^{80}$. Let the SNR of the AWGN channel be $\gamma=0$ [dB] and the target error probability be $\epsilon=10^{-3}$. What is the minimal number of channel uses that needs to be applied? Using numerical solution of (\ref{eq:FIniteChannelUses}) for a complex AWGN, one can find that $n_{\min}=N=128$. We now have to convert this number into a latency figure. However, a channel use is a generic degree of freedom that can carry information. Let the required latency in which this reliability needs to be attained is $T$. Having $N$ and $T$, we can now try to determine how large bandwidth the link needs to use. The number of degrees of freedom that are available in a time-frequency window that spans $T$ seconds and $W$ [Hz] is $2WT$, such that we find the required bandwidth to be:
\begin{equation}
W=\frac{N}{2T}
\end{equation}
Clearly, in order to use the formula (\ref{eq:FIniteChannelUses}), we need to assume that each channel use in the time-frequency grid represents an identical Gaussian channel with $\gamma=0$ [dB]. Nevertheless, our discussion above is an illustration how the finite block length results can lead to latency-constrained transmission techniques. We also note that when the bandwidth is limited to be $W_{\max} < \frac{N}{2T}$, then the required channel uses cannot be obtained in frequency and a possible solution is to e.g. use spatial (MIMO) degrees of freedom. 

Recalling the discussion of the coding of data and metadata, it should be noted that the $N$ channel uses calculated above should contain both the data and the metadata. This implies that the receiver Bob should know in advance that Alice may transmit an ultra-reliable message over the time-frequency grid having $N$ degrees of freedom; only after decoding the message, Bob can verify that it had been intended to him. Keeping the receiver ready over a large bandwidth may not be energy efficient, but this is the investment that Bob can make as a receiver towards achieving URC. On the other hand, Alice can invest a larger transmission power. A good URC system design should strike a good balance between the investments of the transmitter and the receiver.


\section{Wireless Reliability Impairments}

The second dimension for analyzing URC is the type of reliability impairment (RI). We have identified five RIs. 

\subsubsection{Decreased power of the useful signal} This RI refers to the basic propagation mechanisms, such as fading and shadowing. Knowing the statistics of the received signal in the target scenario leads to a proper selection of the coding/modulation parameters for the metadata (e.g. frame synchronization sequence, preambles) and the data. With limited transmission power, the key mechanisms for mitigating this impairment are joint data/metadata encoding, flexible use of the degrees of freedom in frequency and space as well as the new coding techniques for short blocklength. Furthermore, sending reliable short messages over channels with fast dynamics, where the channel estimation at the receiver may not be feasible, require methods for noncoherent communication.

\subsubsection{Uncontrollable interference} This impairment has been the crux of regulating frequency bands. The open access in the unlicensed bands requires to deal with uncontrollable interference, while the high price for a licensed band offers the right to have control over the interference. Nevertheless, the 5G networks will feature sources of unpredictable interference even in the licensed bands. Two examples are ultra-dense deployments of small cells with limited coordination and underlay D2D communication. This RI can be addressed through dynamic spectrum usage, ad hoc cooperation among the interferers, etc.  

\subsubsection{Resource depletion due to competition}
This is in a way similar to the second RI; however, this RI refers to the problem in which multiple devices are trying to share the communication resources in the same system. For example, in reliable coordination among vehicles, each vehicle tries to communicate with all other vehicles, such that they are competing for the same wireless resources. This is the case where resource depletion happens in D2D communication. Traditionally, localized D2D connections have been carried out in unlicensed spectrum. Wireless 5G systems will feature network-controlled D2D communication, where the localized competition for resources among the devices is made more efficient by relying on arbitration and coordination from the cellular network. Network-arbitrated resource competition is one of the key enablers of URC among proximate devices. 

Besides D2D, resource depletion can happen in the downlink (DL) and uplink (UL). In DL the infrastructure has a complete control over the allocation of resources and it can reach the allocation limit if too many devices need to be served. For example, if the number of users in a given area suddenly increases (e.g. public event), then in order to attain the URC-L operation on Fig.~\ref{fig:URCLnrusers}, the signaling in the system needs to have the required level of flexibility and granularity in allocating the resources in order to keep all the users connected. 
In the UL the problem is even more aggravated, due to the lack of coordination across the devices and resource wastes due to collisions, back-off, etc. The key enablers of efficient competition for UL radio resources are \emph{non-orthogonal} operation and successive interference cancellation, as in protocols for coded random access~\cite{Frameless}. 
 
\subsubsection{Protocol reliability mismatch}

The fourth RI refers to the fact that the protocol may be not sufficiently adaptable to offer the required reliability. As discussed in Section~\ref{sec:AnatomyData}, under deteriorating receiving conditions, it becomes a problem to receive the metadata, which is a precondition to receive the data. This RI can be addressed by having protocols that can adapt the transmission of the metadata to the current conditions. We have experimentally shown that such an approach can offer very robust link even with a slight modification of the protocol and without introducing changes in the physical layer~\cite{pcoding}. 

\subsubsection{Equipment failure}

Equipment failure is a RI that is primarily related to disaster/emergency scenarios, where part of the infrastructure becomes dysfunctional. It is addressed through techniques from ad hoc networking, use of D2D communication, etc. 

\section{Conclusion}

Ultra-reliable communication (URC) will be one of the new operating features that will be brought up by the 5G wireless systems. We have provided several motivating scenarios for supporting URC in future wireless applications. We have analyzed the anatomy of a wireless digital link and shown that the introduction of URC requires fundamental rethinking in the relationship between the control information (metadata) and the actual data, since at high reliability levels the way the metadata is encoded and sent cannot be based on the usual ``worst case'' analysis. The paper introduces the important concept of Reliable Service Composition, where a service is designed to adapt its requirements to the level of reliability that can be attained. For example, a service can have a ``minimal variant'' that contains messages that can be encoded and transmitted with very high reliability. We have also introduced different types of URC, long- and short-term, respectively, based on the time frame that is used as a reference to determine the latency of the reliable transmission. Finally, we have identified five general types of reliability impairments that need to be carefully modeled if the system is designed to attain ultra-high reliability levels. 

\section*{Acknowledgement}
Part of this work has been performed in the framework of the FP7 project ICT-317669 METIS, which is partly funded by the European Union. The author would like to acknowledge the contributions of their colleagues in METIS, although the views expressed are those of the authors and do not necessarily represent the project.

\end{document}